\documentclass[aps,twocolumn,showpacs,
superscriptaddress,groupedaddress,floatfix,
twocolumn,aps,prd,amsmath,amssymb]{revtex4}
\usepackage{epsfig}
\usepackage{multirow}
\usepackage{graphicx}
\usepackage{graphics}
\usepackage{subfigure}
\usepackage{epstopdf}
\usepackage{float}
\usepackage{bm}
\usepackage{bbm}
\usepackage{braket}
\usepackage{lineno}

\begin{document}

\title{Renormalization of the band gap in 2D materials through the competition between electromagnetic and four-fermion interactions}

\author{Luis Fern\'andez}
\email{luis.aguilar@icen.ufpa.br} \affiliation{Faculdade de F\'\i sica, Universidade Federal do Par\'a, 66075-110, Bel\'em, PA,  Brazil}

\author{Van S\'ergio Alves}
\email{vansergi@ufpa.br}
\affiliation{Faculdade de F\'\i sica, Universidade Federal do Par\'a, 66075-110, Bel\'em, PA,  Brazil}

\author{Leandro O. Nascimento}
\email{lon@ufpa.br}
\affiliation{Faculdade de Ci\^encias Naturais, Universidade Federal do Par\'a, C.P. 68800-000, Breves, PA,  Brazil}

\author{Francisco Pe\~na}
\email{francisco.pena@ufrontera.cl} 
\affiliation{Departamento de Ciencias F\'\i sicas, Facultad de Ingenier\'\i a y Ciencias, Universidad de La Frontera, Avda. Francisco Salazar 01145, Casilla 54-D, Temuco, Chile}

\author{M. Gomes}
\email{mgomes@fma.if.usp.br} 
\affiliation{Instituto de F\'\i sica, Universidade de S\~ao Paulo Caixa Postal 66318, 05315-970, S\~ao Paulo, SP, Brazil}

\author{E. C. Marino}
\email{marino@if.ufrj br}
\affiliation{Instituto de F\'\i sica, Universidade Federal do Rio de Janeiro, 21941-972, Rio de Janeiro, RJ, Brazil}

\date{\today}

\begin{abstract}

Recently the renormalization of the band gap $m$, in both WSe$_2$ and MoS$_2$, has been experimentally measured as a function of the carrier concentration $n$. The main result establishes a decreasing of hundreds of meV, in comparison with the bare band gap, as the carrier concentration increases. These materials are known as transition metal dichalcogenides and their low-energy excitations are, approximately, described by the massive Dirac equation. Using Pseudo Quantum Electrodynamics (PQED) to describe the electromagnetic interaction between these quasiparticles and from renormalization group analysis, we obtain that the renormalized mass describes the band gap renormalization with a function given by $m(n)/m_0=(n/n_0)^{C_\lambda/2}$, where $m_0=m(n_0)$ and $C_\lambda$ is a function of the coupling constant $\lambda$. We compare our theoretical results with the experimental findings for WSe$_2$ and MoS$_2$, and we conclude that our approach is in agreement with these experimental results for reasonable values of $\lambda$. In addition we introduced a Gross-Neveu (GN) interaction which could simulate an disorder/impurity-like microscopic interaction. In this case, we show that there exists a critical coupling constant, namely, $\lambda_c \approx 0,66$ in which the beta function of the mass vanishes, providing a stable fixed point in the ultraviolet limit. For $\lambda>\lambda_c$, the renormalized mass decreases while for $\lambda<\lambda_c$ it increases with the carrier concentration.  


\end{abstract}

\pacs{11.10.Hi, 11.15.-q, 11.15.Pg, 71.10.Pm}

\maketitle
\section{Introduction}

The interest in two-dimensional materials has been increased due to several new applications, in particular, the control of charge, spin, and valley of electrons in the honeycomb lattice. The understanding of the material properties and its fundamental interactions have been discussed in several experimental and theoretical studies, aiming to applications  and development of electronic devices with these materials, in particular, for graphene \cite{grapexp}, silicene \cite{silexp}, and transition metal dichalcogenides \cite{TMDs}.

Although a full description of the material properties would require the inclusion of several microscopic interactions, such as the lattice and impurities, it is possible to focus on a low-energy description of the electrons close to the Dirac points, also called the valleys of the honeycomb lattice. In this case, a quantum-electrodynamical approach is derived which is expected to describe electronic properties at low temperatures \cite{Voz1, PRX, Exc}. Within this regime, electrons obey a Dirac-like equation with two main parameters, namely, the Fermi velocity $v_F$ and the mass (band gap) $m$. In graphene, for instance, the energy gap vanishes and the Fermi velocity reads $v_F\approx c/300$ where $c$ is the light velocity. In other materials, like silicene and TMDs, the breaking of sublattice symmetry yields a nonzero energy gap of the order 1-2 eV. Effects of interactions, nevertheless, may renormalize these quantities yielding results that are dependent on the coupling constants and the electron density, as it has been shown in the case of the renormalization of the Fermi velocity in clean graphene due to a static Coulomb potential . 

As it is well-known, the renormalization of $v_F$ in graphene shows that, at very low densities, one finds an ultra-relativistic regime where $v_F \rightarrow c$, recovering the so-called Lorentz symmetry \cite{geim}. Despite the experimental difficulty to actually reach this regime, it is remarkable that the effect of interactions yields a possible realization of massless Dirac fermions in a two-dimensional crystal. It is worth to mention that PQED \cite{marino2} applied to graphene yields a suitable description for both cases either for $v_F \ll c$ or $v_F \approx c$ \cite{libroMarino}. Therefore, the renormalization of $v_F$ is straightforward within this quantum-electrodynamical approach. Indeed, in similar systems, the use of quantum field theory techniques has been shown very useful for describing  electronic properties \cite{Exc,PRX,Bfraco,Placa,Cavidade,Gfactor,Quiral, Dual}. Nevertheless, the effects of GN interactions coupled to PQED have been less discussed \cite{PQEDGNSD}.

In this work, we shall investigate the effect of electromagnetic interaction on the quasiparticle mass renormalization of two-dimensional systems via analysis of the renormalization group in the dominant order at 1/N. We shall use PQED, sometimes called reduced quantum electrodynamics \cite{Teber1}, to describe the interaction between these quasiparticles through the Gauge field. We compare our theoretical results with the recent experimental findings for WSe$_2$ \cite{WSe2gap} and MoS$_2$ \cite{MoS2gap}. Next, using GN-type interaction to simulate some impurity/disorder present in the sample of a 2D-Dirac material \cite{fermisys}, we shall investigate the influence of this interaction on the PQED renormalization group functions.  

This paper is organized as follow. In Sec.~II, we present the model, Feynman's rules and obtain photon propagator and electron self-energy both in the dominant order of 1/N. In Sec.~III, we analyze the renormalization group functions and we obtain the renormalized mass which describes the band gap renormalization and we compare our theoretical results with the experimental findings for WSe$_2$ and MoS$_2$. In Sec.~IV we investigated the influence of GN interaction in large-N expansion in the renormalization group functions obtained in the previous section. In Sec.~V, we review the main results obtained in this paper and in the App.~A-C we show some details of the calculations.

\section{Electromagnetic Interactions for Massive Electrons in Two-Dimensions}

In this section, we calculate some effects of the electromagnetic interactions for two-dimensional materials with a band gap. This band gap may be described as a mass term at the level of Dirac equation. This is a consequence of the tight-binding approximation for electrons in the honeycomb lattice at low energies (See Ref.~\cite{Exc} for a full derivation in the supplementary material). 

Let us consider a Lagrangian in Euclidean space given by
\begin{equation}
\begin{split}
\mathcal{L}=&\frac{1}{2}\frac{F^{\mu \nu}F_{\mu \nu}}{(-\Box)^{\frac{1}{2}}} + \dot{\imath}\bar{\psi}_a\left(\gamma^0\partial_0+v_F\gamma^{i}\partial_i-m\right)\psi_a  \\
&-\frac{\xi}{2}A_{\mu}\frac{\partial^{\mu}\partial^{\nu}}{\left(-\Box\right)^{\frac{1}{2}}}A_{\nu}+e\bar{\psi}_a\gamma^{\mu}\psi_a A_{\mu}, 
\label{L0}
\end{split}
\end{equation}
where $v_F$ is the bare Fermi velocity, $e$ is the electromagnetic coupling constant, and $m$ is the bare mass of the electron. $A_{\mu}$ is the pseudo-electromagnetic field and $F_{\mu\nu}$ is its usual field-intensity tensor. $\psi_a$ is the Dirac field describing the electrons of the p-orbitals in the honeycomb lattice that are relevant for describing electronic properties. Furthermore, $a=1,...,N$ is a flavor label for this matter field that aims for describing both valley and spins indexes (or any other internal symmetry). Here, our matter field reads $\psi_a^{\dagger}=(\psi^{*}_{A \uparrow}, \psi^{*}_{A\downarrow}, \psi^{*}_{B\uparrow}, \psi^{*}_{B\downarrow})_a$, where $(A,B)$ and $(\uparrow,\downarrow)$ are the sublattices and spin of the honeycomb lattice, respectively. Therefore, $a=K, K'$ and $N=2$ describe the valley degeneracy \cite{Gfactor, libroMarino}. $\xi$ is the Gauge fixing parameter. From Eq.~(\ref{L0}), we obtain the energy dispersion $E_\pm(\textbf{p})=\pm \sqrt{v^2_F \textbf{p}+m^2}$, where the sign $\pm$ means either the valence band ($-$) or the conduction band ($+$). Note that we are using the natural system of units, where $\hbar=c=1$.

We may conclude from the Dirac Lagrangian in Eq. \eqref{L0} that the characteristic exponent of the space-time anisotropy is given by $z = 1$. This is given by the exponent of the higher-order derivative term that breaks Lorentz symmetry, i.e, $(v_F\partial_i )^z$. Therefore, as it
has been shown in Ref. \cite{gomes2}, this means a soft breaking of the Lorentz symmetry. Higher-order terms would imply higher-order derivatives, which, in principle, could describe the behavior of electrons far from the Dirac point.

We consider the large-$N$ expansion at one-loop approximation and in this case of a trilinear interaction like that of Eq.\eqref{L0} this can be done through the substitution $e\rightarrow e/\sqrt{N}$, for fixed $e$. Thus, the Feynman rules, in the Euclidean space,  are
\begin{equation}
S_F(p,m) =  \frac{\gamma^0 p_0 + v_F \gamma^i p_i + m}{p_0^{2}+v_F^{2}\textbf{p}^2+m^2}, 
\label{freefermion}
\end{equation}
which is the Fermion propagator,
 \begin{equation}
 \Delta_{\mu \nu}^{(0)}(p)=\frac{1}{2 \epsilon \sqrt{p^2}}\left[\delta_{\mu \nu}-\left(1-\frac{1}{\xi}\right)\frac{p_{\mu} p_{\nu}}{p^2}\right], \label{nufot} 
 \end{equation}
for the Gauge-field propagator, where $\epsilon$ is the dielectric constant of the medium, and
\begin{equation}
\frac{e}{\sqrt{N}}\gamma^{\mu}, \mbox{and}\,\gamma^{\mu} \rightarrow \left\{ \begin{array}{lc}
             \mu = 0, & \gamma^0    \\
             \\ \mu = i, &  v_F \gamma^i
             \end{array}
   \right.,
\end{equation}
describing the electromagnetic interaction.

\subsection{The Gauge-field propagator}

\begin{figure}[H]
\centering
\includegraphics[scale=0.5]{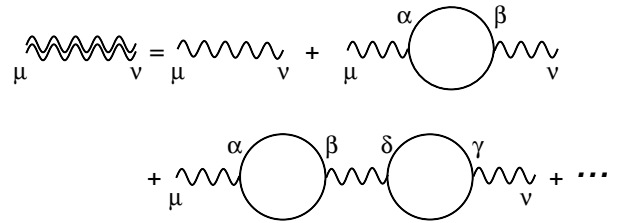}
\caption{\textit{The full propagator of Gauge field in the dominant order of $1/N$.}}
\label{loops}
\end{figure}

The full propagator of the Gauge field is calculated, in momentum space, from
\begin{equation}
\Delta_{\mu \nu}(p)=\Delta_{\mu \nu}^{(0)}(p)+\Delta_{\mu \alpha}^{(0)}(p)\Pi^{\alpha \beta}(p)\Delta_{\beta \nu}^{(0)}(p)+ \cdots, 
\label{geoloops}
\end{equation}
which is a geometric series. In the large-$N$ approximation, the quantum corrections may be expressed as a sum over diagrams of the same order in the parameter $N$, as it is shown in Fig. \ref{loops}, since $e\rightarrow e/\sqrt{N}$.
The polarization tensor is
\begin{equation}
\Pi^{\mu \nu}(p)=-e^2\,{\rm Tr} \int \frac{d^3 k}{(2\pi)^3} \gamma^{\mu} S_F(k)\gamma^{\nu}S_F(k+p).
\label{polk}
\end{equation}

Next, we assume that the interaction vertex is just given by $\gamma^{0}$ (this means we are assuming a static regime). Using  the dimensional regularization we obtain the time component of the polarization tensor, which is given by (See App.~A for more details)
\begin{equation}
\Pi^{00}(p)=-\frac{e^2 }{8}\left[\frac{\textbf{p}^2}{\sqrt{p_0^{2}+v_F^{2}\textbf{p}^{2}}}-\frac{4\textbf{p}^2 m^{2}}{(p_0^{2}+v_F^{2}\textbf{p}^{2})^{\frac{3}{2}}}\right].
\label{pi}
\end{equation}
Thereafter, we use Eq.~\eqref{pi} and the free photon propagator, given in Eq. \eqref{nufot}, for calculating the full propagator of the Gauge field. This is given by
\begin{equation}
\begin{split}
\Delta_{00}(p)&=\left(2\epsilon\sqrt{\textbf{p}^2}+\frac{e^2 }{8}\left[\frac{\textbf{p}^2}{\sqrt{p_0^{2}+v_F^{2}\textbf{p}^{2}}}\right. \right. \\& \left. \left.-\frac{4\textbf{p}^2 m^{2}}{(p_0^{2}+v_F^{2}\textbf{p}^{2})^{\frac{3}{2}}}\right]\right)^{-1},
\label{fullfot}
\end{split}
\end{equation}
where the static approximation has been implemented, which consists of taking $p_0=0$ at the free photon propagator.

\subsection{The fermion self-energy}

\begin{figure}[H]
\center
\includegraphics[scale=0.55]{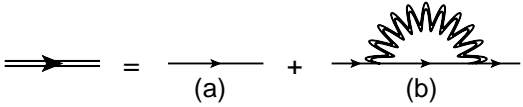}
\caption{\textit{The full fermion propagator}. $ (a) $ represents the free fermion propagator, $(b)$ is the 1-loop correction due to the full photon propagator in the dominant order $1/N$.}
\label{fermifullPQED}
\end{figure}

The fermion propagator with the self-energy corrections, in the dominant order of 1/N, is shown in Fig. \ref{fermifullPQED}.

Let us first calculate the self-energy of the fermion due to the electromagnetic interaction, represented in Fig. \ref{fermifullPQED} $(b)$. Using the static approximation, the electron self-energy reads
\begin{equation}
\Sigma_{A_{\mu}}(p)= \frac{e^2}{N}\int \frac{d^3 k}{(2 \pi)^3} \gamma^0 S_F(p-k)\gamma^0 \Delta_{00}(k).
\label{sigmaa2}
\end{equation}
Because we are interested in the small momentum behavior of this expression (similar to the approximation used in Ref.~\cite{son}), we can write
\begin{equation}
\Sigma_{A_{\mu}}(p)= \Sigma_{A_{\mu}}^{(0)} + \gamma^0p_0\Sigma_{A_{\mu}}^{(1a)} +v_F\gamma^ip_i\Sigma_{A_{\mu}}^{(1b)},
\label{sigmaa}
\end{equation}
where $\Sigma_{A_{\mu}}^{(0)}$, $\Sigma_{A_{\mu}}^{(1a)}$, and $\Sigma_{A_{\mu}}^{(1b)}$ are the lowest-order terms with
\begin{equation}
\Sigma_{A_{\mu}}^{(0)}=-\frac{e^2}{N}\int \frac{d^3 k}{(2\pi)^3}\left[\frac{m}{k_0^{2}+v_F^{2}\textbf{k}^2+m^{2}}\right]\Delta_{00}(k),
\label{f0}
\end{equation}
\begin{equation}
\Sigma_{A_{\mu}}^{(1a)}\!=\!-\frac{e^2}{N}\int \!\!\frac{d^3 k}{(2\pi)^3}\left[\frac{v_F^{2}\textbf{k}^2-k_0^{2}+m^{2}}{(k_0^{2}+v_F^{2}\textbf{k}^2+m^{2})^2}\right]\Delta_{00}(k),
\label{f1}
\end{equation}
and
\begin{equation}
\Sigma_{A_{\mu}}^{(1b)}=\frac{e^2}{N}\int \frac{d^3 k}{(2\pi)^3}\left[\frac{m^{2}+k_0^{2}}{(k_0^{2}+v_F^{2}\textbf{k}^2+m^{2})^2}\right]\Delta_{00}(k).
\label{f2}
\end{equation}
 
Next, we perform a variable change $v_F k_i \rightarrow \overline{k}_i$ and, using spherical coordinates, the full photon propagator can be written as
\begin{equation}
\Delta_{00}(k)=\!\!\frac{v_F}{2 \epsilon}\frac{1}{k \sin\theta}\!\!\left[1\!+\!\frac{e^{2}}{16 \epsilon v_F}\!\!\left(1-\frac{4 m^{2}}{k^2}\right)\sin\theta\right]^{-1}.
\label{fotonfull}
\end{equation}
Using the small-mass limit ($m^2\ll k^2$) the term $4m^2/k^2$ can be neglected in Eq.~(\ref{fotonfull}). Furthermore, since we are studying the model in the static approximation, with $\lambda= e^2/16\epsilon v_F<1$, hence, we have 
\begin{eqnarray}
\Sigma_{A_{\mu}}(p)&=&-\frac{2 \lambda}{\pi^2 N} [m f_0(\lambda) + \gamma^0 p_0 f_1(\lambda)  \nonumber\\ 
&-& v_F \gamma^i p_i f_2(\lambda)]\ln\left(\frac{\Lambda}{\Lambda_0}\right)+ \textrm{FT},
\label{autofoton}
\end{eqnarray}
where ${\rm FT}$ stands for finite terms, $\Lambda$ and $\Lambda_0$ are ultraviolet and infrared cutoff respectively. The functions $f_0(\lambda)$, $f_1(\lambda)$, and $f_2(\lambda)$ are given by (See App.~B for more details and Ref.\cite{son})
\begin{equation}
f_0(\lambda)= \frac{2\cos^{-1}(\lambda)}{\sqrt{1-\lambda^2}},\\
\label{f0}
\end{equation} 
\begin{equation}
f_1(\lambda)=-\frac{2}{\lambda^2}\left[\pi-2\lambda+\frac{(\lambda^2-2)}{\sqrt{1-\lambda^2}}\cos^{-1}(\lambda)\right],\\
\label{f1}
\end{equation} 
and
\begin{equation}
f_2(\lambda)=\frac{1}{\lambda^2}\left[\pi-2\lambda-2\sqrt{1-\lambda^2}\cos^{-1}(\lambda)\right].
\label{f2}
\end{equation} 

\section{Renormalization group}

In principle the renormalization group (RG) equation presents two anomalous dimensions corresponding to each field $\psi$ and $A_{\mu}$. However, since the polarization tensor for the Gauge is finite, within the dimensional regularization, we may conclude that $\gamma_{A_{\mu}}=0$, and, therefore, $\beta_e=0$. Hence, the RG equation reads  
\begin{equation}
\!\!\left[\Lambda \frac{\partial}{\partial \Lambda}+\beta_{v_F}\frac{\partial}{\partial v_F}+\beta_{m}\frac{\partial}{\partial m}- N_F \gamma_F\right]\Gamma^{(N_F,N_A)}(p_i)=0,
\label{GRPQED}
\end{equation}
where $\Gamma^{(N_F,N_A)}(p_i=p_1,...,p_N)$ means the renormalized vertex functions. $N_F$ and $N_A$ are the number of external lines of fermion and Gauge fields, respectively. $\beta_{v_F}=\Lambda\frac{\partial v_F}{\partial\Lambda}$ and $\beta_{m}=\Lambda\frac{\partial m}{\partial\Lambda}$ are the beta functions of the parameters $v_F$ and $m$, respectively. The function $\gamma_F$ is the anomalous dimension of the fermion, given by $\gamma_F=\Lambda\frac{\partial}{\partial\Lambda}\left(\ln\sqrt{Z_{\psi}}\right)$, where $Z_{\psi}$ is the wavefunction renormalization. For our purpose, it is sufficient to consider only the vertex function for the fermion, i.e, $\Gamma^{(2,0)}$. Therefore, we can write 
\begin{equation}
\Gamma^{(2,0)}= \left(\gamma^0 p_0 + v_F\gamma^i p_i - m\right) + \Sigma_{A_{\mu}}(p) . \label{gammavertexPQED}
\end{equation}

Now we must replace Eq. \eqref{gammavertexPQED} in Eq.~(\ref{GRPQED}). Note that, despite our notation, the parameters $v_F$ and $m$ inside Eq. \eqref{gammavertexPQED} are the renormalized parameters in agreement with Eq.\eqref{GRPQED}. We write the betas functions as a series, such that  $\beta_a=N^0\,\beta_a^{(0)}+\frac{1}{N}\,\beta_a^{(1)}+...$ for $a=v_F,m$, and $\gamma_F=N^0\,\gamma_F^{(0)}+\frac{1}{N}\,\gamma_F^{(1)}+...$, thus, we can write  
\begin{equation}
\beta_{v_{F}}=-\frac{4}{\pi^2 N}v_F\left[1+\frac{\cos^{-1}(\lambda)}{\lambda\sqrt{1-\lambda^2}}\right]+\frac{2}{\pi N}\frac{v_F}{\lambda}, \label{betavelocidad0}
\end{equation}

\begin{equation}
\beta_{m}=-\frac{2}{\pi^2 N}m\left[4 +\frac{4 \cos^{-1}(\lambda)}{\lambda\sqrt{1-\lambda^2}}-\frac{2\pi}{\lambda}\right],
\label{betamass0}
\end{equation}
and for the anomalous dimension we have
\begin{eqnarray}
\gamma_F&=&-\frac{2}{\pi^2 N}\left[2+\frac{2-\lambda^2}{\lambda\sqrt{1-\lambda^2}}\cos^{-1}(\lambda)\right]\nonumber\\ &+&\frac{2}{\pi N}\frac{1}{\lambda}. \label{anomala0}
\end{eqnarray}
The coupling constant $\lambda$ is defined as $\lambda\equiv e^2/16 \epsilon v_F=\pi \alpha/4$, where $\alpha$ is the dimensionless fine-structure constant. 

\subsection{Fermi velocity renormalization and anomalous dimension}

Using the definition of the beta function for $v_F$, namely, $\beta_{v_F}=\Lambda\frac{\partial v_F}{\partial \Lambda}$ in Eq. \eqref{betavelocidad0}, we may find the renormalized Fermi velocity depending on the energy scale $\Lambda$. We may replace the energy scale by the carrier concentration $n$ by performing the following transform $\Lambda/\Lambda_0 \rightarrow (n/n_0)^{1/2}$. After doing this, it has been shown that Eq. \eqref{betavelocidad0}, with an effective dielectric constant, yields a very good agreement with the experimental data for graphene \cite{geim, Gui}. The main effect is that the value of $v_F$ increases as we decrease the value of $n$. This may be improved by producing more and more clean samples. Here, we conclude that the presence of the mass \textit{does not change} this result. Therefore, a similar renormalization of the Fermi velocity, in other two-dimensional materials, is expected to occur.

Eq. \eqref{anomala0} yields the anomalous dimension of the model at one-loop approximation. Note that for a very large number of fermionic species, $N\rightarrow \infty$, the anomalous dimension vanishes.
\begin{figure}[H]
\includegraphics[scale=0.65]{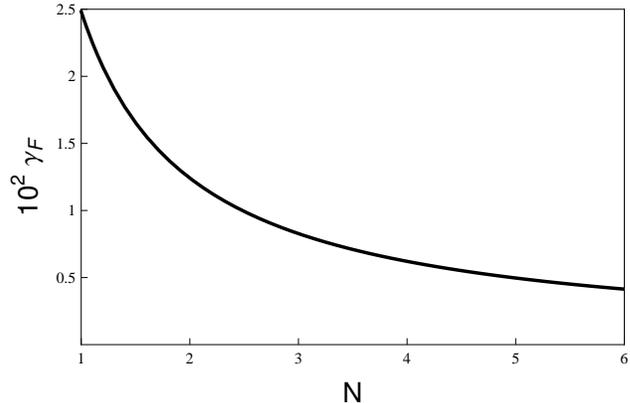}
\caption{\textit{Anomalous Dimension}. We plot Eq.~(\ref{anomala0}) with $\lambda=0.9$ as a function of $N$. In the asymptotic limit $N\rightarrow \infty$, $\gamma_F\rightarrow 0$ and the fermion field recovers its classical dimension.}
\label{gamma}
\end{figure}

\subsection{Mass Renormalization}

\begin{figure}[H]
\centering
\includegraphics[scale=0.65]{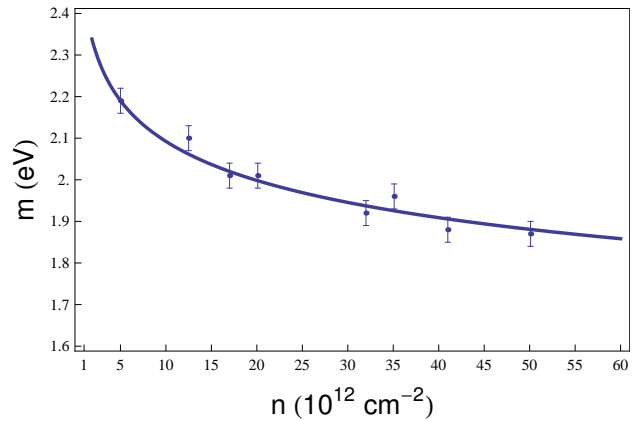}
\caption{\textit{Renormalization of the band gap for MoS$_2$}. The thick line is the plot of Eq.~(\ref{sigR0}) with $C_\lambda=-0.13$ ($\lambda \approx 0.63$) and $m_0=2.19$eV at $n_0=5.01 \times 10^{12}$cm$^{-2}$. We have assumed $N=2$ (spin degeneracy). The small dots and error bars have been extracted from Fig.~4 in Ref.~\cite{MoS2gap}. They are the experimental values of the renormalized band gap at different values of the carrier concentration $n$ for MoS$_2$ at $T=$295K.} 
\label{comparison3}
\end{figure}

Using the definition of the beta function for $m$, namely, $\beta_{m}=\Lambda\frac{\partial m}{\partial \Lambda}$ in Eq. \eqref{betamass0}, we may find the renormalized mass depending on the energy scale $\Lambda$. This, nevertheless, may be described in terms of $n$ as discussed before. After a simple calculation, we find
\begin{equation}
m(n)=m_0\left(\frac{n}{n_0}\right)^{C_{\lambda}/2}, \label{sigR0}
\end{equation}
where $m_0\equiv m(n_0)$ is a reference value (which must be provided by experiments) and
\begin{equation}
C_\lambda=-\frac{2}{\pi^2 N}\left[4+\frac{4 \cos^{-1}(\lambda)}{\lambda\sqrt{1-\lambda^2}}-\frac{2\pi}{\lambda}\right] \label{Cl}
\end{equation}
is a known constant fixed by the coupling constant $\lambda$ and $N=2$. Within the realm of two-dimensional materials, Eq.~(\ref{sigR0}) shows that the band gap is tunable by changing the carrier concentration $n$. The renormalization of $m$ has been experimentally measured in Ref.~\cite{WSe2gap}, where the authors have shown that, by changing the carrier concentration from $n\approx 1.6 \times 10^{12}$cm$^{-2}$ to $n\approx 1.5 \times 10^{13}$cm$^{-2}$, the energy gap decreases approximately 400meV of its bare value for tungsten diselenide WSe$_2$. Beyond several kind of applications, this effect could be useful for studying the electric-field tuning of energy bands with nontrivial topological properties.
In Ref.~\cite{MoS2gap}, a similar result has been found for Molybdenum disulfide MoS$_2$. 
\begin{figure}[H]
\includegraphics[scale=0.65]{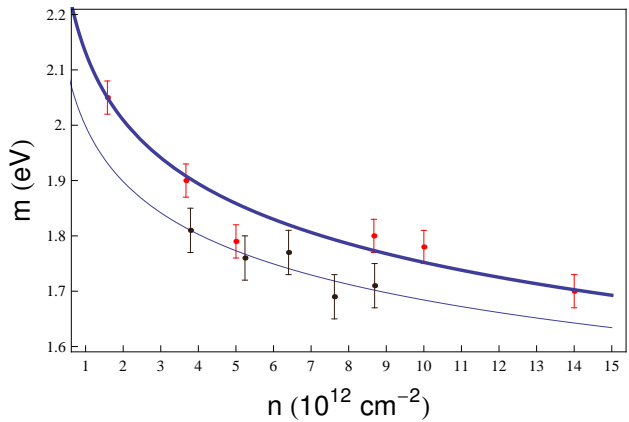}
\caption{\textit{Renormalization of the band gap for WSe$_2$}. The thick line is the plot of Eq.~(\ref{sigR0}) with $C_\lambda=-0.17$ ($\lambda \approx 0.96$) and $m_0=2.05$eV at $n_0=1.58 \times 10^{12}$cm$^{-2}$. The common line is the plot of Eq.~(\ref{sigR0}) with $C_\lambda=-0.15$ ($\lambda \approx 0.76$) and $m_0=1.81$eV at $n_0=3.79 \times 10^{12}$cm$^{-2}$. For both curves we have assumed $N=2$ (spin degeneracy). The small dots and error bars have been extracted from Fig.~4 in Ref.~\cite{WSe2gap}. They are the experimental values of the renormalized band gap at different values of the carrier concentration $n$ for WSe$_2$ at $T=$100K. The two different colors are related to the two different devices with different thickness of the boron nitride substrate. The red point is device 1 with $d_{{\rm BN}}\approx 7.4$nm and the black point is device 2 with $d_{{\rm BN}}\approx 4.5$nm. This implies that there exist two different $\lambda$ constants in our theoretical result in Eq.~(\ref{sigR0}).} 
\label{comparison}
\end{figure}
In FIG.~\ref{comparison}, we compare our analytical result with the experimental data for WSe$_2$, using the corresponding error bars for each point \cite{WSe2gap}. This experimental data has been obtained by putting the monolayer of WSe$_2$ above two different substrates made of boron nitride. It is well-known that the substrate changes the fine-structure constant $\alpha$, because of its effects on the dielectric constant $\epsilon$. Within our result, in Eq.~(\ref{sigR0}), this may be described by considering a different value of the constant $\lambda=\pi \alpha/4$ for each substrate. Furthermore, we choose a reference value of $m_0=2.05$eV at $n_0=1.58 \times 10^{12}$cm$^{-2}$, which plays the role of ``bare" mass for our purposes. This is a mandatory step in the renormalization procedure in general. Thereafter, we choose a best fitting parameter $C_\lambda$ that, essentially, provides the value of the coupling constant $\lambda$. Using these results, we conclude that the expected values for $\alpha$ are between $\alpha \approx [0.97, 1.22]$. This estimative is close to the result used in Ref.~\cite{Exc} for describing the formation of excitons (pairs of electrons and holes), where $\alpha \approx 0.65$ (the small difference is due to the use of different substrates for each device). We obtain that the decreasing of the energy gap, whether $n\rightarrow 1.5\times 10^{13}$cm$^{-2}$, is of the order of 400meV in comparison with the bare gap, as expected \cite{WSe2gap}. Furthermore, besides one point (red point below the thick line in Fig.~\ref{comparison}), the other experimental data are within our theoretical result. In particular, all of the black points are described by our result. In Fig.~\ref{comparison3}, we repeat the same procedure, but for MoS$_2$. In this case, the experimental data are within our theoretical result in Eq.~(\ref{sigR0}).   

Although small deviations are expected to occur, because the experimental data has been obtained at temperatures of 100K, nevertheless, this effect is small. Indeed, note that the activation temperature $T^*$ from the minimum of the valence band to the minimum of the conduction band is of the order of the bare gap, i.e, $\approx 1$eV, which means $T^* \approx 10^4$K roughly speaking. Hence, the effects of the thermal bath are relevant for temperatures close to $10^4$K, far beyond the room temperature. We believe that either higher-order corrections or inclusion of more interactions at one-loop level are likely to improve this comparison. The measurement of more experimental points are also relevant for a more precise comparison. It is worth to mention that both materials, WSe$_2$ and MoS$_2$, have an excitonic spectrum that has been accurately described by PQED in Eq.~(\ref{L0}), in particular, with a good agreement with the experimental findings \cite{Exc}.

\section{Effects of Four-Fermion Interactions}
To include a four-fermion interaction in our model given by Eq.~(\ref{L0}), we star, just for the fermion sector, the following Lagrangian 
\begin{eqnarray}
\mathcal{L}&=& \dot{\imath}\bar{\psi}_a\left(\gamma^0\partial_0+v_F\gamma^{i}\partial_i\right)\psi_a  - \frac{g_0}{2}\left(\bar{\psi}_a\psi_a\right)^{2},
\label{Lmod}
\end{eqnarray}
where $g$ is the bare coupling constant of the self-interaction between fermions. Because the GN interaction can be related to the disorder/impurity, which modifies the density of states in monolayer graphene \cite{den1,den2}, a more realistic model for describing transport properties in these systems should include four-fermions interactions. Note that although we have performed $m=0$ in Eq.~(\ref{Lmod}), we know that the GN interaction generates a mass for the fermion in 1/N expansion by the chiral symmetry breaking. Also, it well-known that the GN interaction \cite{gross} is non-renormalizable in the coupling constant $g$. Nevertheless, it is renormalizable in the context of the  1/N expansion \cite{gross}. In this expansion, we must perform the following transformations  $g_ \rightarrow \frac{g}{N}$, for fixed $g$.

Next, we introduce an auxiliary field in Eq. \eqref{Lmod} in order to convert the Gross-Neveu interaction into a trilinear one. This auxiliary field may represent the presence of  disorder/impurities \cite{wang, fermisys} in the  model. Hence,
\begin{align}
\mathcal{L}& = \mathcal{L}+\frac{N}{2g}\left(\sigma - \frac{g}{N}\bar{\psi_a}\psi_a\right)^2 \nonumber\\
&=\mathcal{L}+\frac{N}{2g}\sigma^2-\sigma\bar{\psi_a}\psi_a. \label{auxfield}
\end{align}
This new field $\sigma$ does not change the dynamics of system, because it represents only a constrained that is derived from the Euler-Lagrange equation as
\begin{equation}
\sigma=\frac{g}{N}\bar{\psi_a}\psi_a.
\end{equation}

Eq. \eqref{auxfield} shows that a mass term is generated for the fermion field whether $\langle \sigma \rangle\neq 0$. This implies a spontaneous breaking of the discrete chiral symmetry $\psi_a \rightarrow \gamma^5 \psi_a$ and we shall be working in this broken phase. This is the reason why we have assumed $m=0$ in Eq. \eqref{Lmod}. Indeed, we may define $\braket{\sigma}= \sigma_0$, hence, $\sigma_0$ represents the mass generated for the electrons \cite{gross}. Thereafter, we replace $\sigma \rightarrow \sigma_0 + \frac{\sigma}{\sqrt{N}}$ for ensure that the theory has a ground state. Hence, the Lagrangian reads
\begin{equation}
\begin{split}
 \mathcal{L}&=\bar{\psi_a}\left(\dot{\imath}\gamma^0\partial_0 + \dot{\imath}v_F\gamma^{i}\partial_i - \sigma_0\right)\psi_a +\frac{N}{2g}\sigma^2_{0} \\ &+ \frac{\sqrt{N}}{g}\sigma_0\sigma + \frac{1}{2g}\sigma^2- \frac{1}{\sqrt{N}}\sigma\bar{\psi_a}\psi_a. 
 \end{split}
 \label{Lmod2}
 \end{equation}
The free propagator of fermion field is given by the eq. \eqref{freefermion} where $m \rightarrow \sigma_0$ and the auxiliary-field propagator reads 
\begin{equation}
\Delta_{\sigma}^{0}=\left(\frac{1}{g}\right)^{-1}, \label{freeaux}
\end{equation}
and the vertex interaction is given by $-\frac{1}{\sqrt{N}}$, describing the GN interaction.

\subsection{The auxiliary-field Propagator}

The quantum correction for the auxiliary-field propagator in the lowest order of $1/N$ can be calculated from the functional integral method. We can rewrite Eq. \eqref{Lmod2} and obtain
\begin{eqnarray}
{\cal L}=\bar\psi_a K\psi_a+\frac{N}{2g}\sigma^2_{0} + \frac{\sqrt{N}}{g}\sigma_0\sigma +\frac{1}{2g}\sigma^2, \label{auxaction}
\end{eqnarray}
where $K=\dot{\imath}\gamma^0\partial_0+\dot{\imath}v_F\gamma^i\partial_i - \sigma_0 - \frac{1}{\sqrt{N}}\sigma$. Integration over $\psi$ in Eq.~(\ref{auxaction}) yields the effective action $S_{{\rm eff}}[\sigma]$ for the auxiliary field, given by $S_{{\rm eff}}= N \,{\rm Tr} \ln K \approx \sqrt{N}S_1+N^0 S_2+...$ , where the last equality has been obtained for large-$N$. Furthermore,
\begin{eqnarray}
S_1&=&-{\rm Tr}\left[\left(\dot{\imath}\gamma^0\partial_0+\dot{\imath}v_F\gamma^i\partial_i-\sigma_0\right)^{-1}\sigma\right] \nonumber\\
&+& \int d^3 x \frac{1}{g}\sigma_0\sigma, \label{s1}
\end{eqnarray}
and
\begin{eqnarray}
S_2&=& {\rm Tr}\left[\left\lbrace\left(\dot{\imath}\gamma^0\partial_0+\dot{\imath}v_F\gamma^i\partial_i-\sigma_0\right)^{-1}\sigma\right\rbrace^2\right] \nonumber\\ 
&&+ \int d^3 x \frac{1}{g}\sigma^2. \label{s2}
\end{eqnarray}

Eq.~\eqref{s1} yields the so-called gap equation after we assume $S_1=0$ in order to have a finite effective action. This gap equation reads
\begin{eqnarray}
\frac{1}{g}&=& 4\int \frac{d^3p}{(2\pi)^3}\frac{1}{p_0^{2}+v_F^{2}\textbf{p}^{2}+\sigma_0^{2}} \nonumber\\
\label{masagap}
\frac{1}{g}&=&-\frac{1}{\pi v_F^{2}}|\sigma_0|.
\end{eqnarray}
This equation shows that we can relate the generated mass $\sigma_0$ with the coupling constant $g$ \cite{Livros,Dimensional}.

Eq.\eqref{s2} may be written as
\begin{equation}
S_2=\frac{1}{2}\int d^3x d^3y \,\sigma(x)\Gamma_{\sigma}(x-y)\sigma(y),
\end{equation} 
where $\Gamma_{\sigma}$ is the inverse of the full Auxiliary-field propagator, hence,
\begin{equation}
\Gamma_{\sigma}(p) = \frac{1}{g}+  {\rm Tr} \int \frac {d^3k}{(2\pi)^3} \left[S_F(k+p)S_F(k) \right]. 
\label{Dsigma}
\end{equation} 
Therefore,
\begin{equation}
\Delta^{-1}_{\sigma}(p)=\Gamma_{\sigma}(p)=(\Delta_{\sigma}^{0})^{-1} + \Pi_{\sigma}(p),
\label{deltaaux}
\end{equation}
where $\Pi_{\sigma}(p)$ is the self-energy due to the Gross-Neveu interaction. Note that Eq.~(\ref{deltaaux}) is the well-known Schwinger-Dyson equation for the $\sigma$ field. Using the Eq. \eqref{masagap} in the Eq. \eqref{Dsigma} and after some simplifications, we find that (see App.~C for more details)
\begin{equation}
\begin{split}
\Pi_{\sigma}(p)&=\frac{1}{\pi v_F^{2}}\left[|\sigma_0|+\frac{p_0^{2}+v_F^{2}\textbf{p}^{2}+4\sigma_0^{2}}{2\sqrt{p_0^{2}+v_F^{2}\textbf{p}^{2}}} \times \right. \\ & \left. \sin^{-1}\left(\sqrt{\frac{p_0^{2}+v_F^{2}\textbf{p}^2}{p_0^{2}+v_F^{2}\textbf{p}^{2}+4\sigma_0^{2}}}\right)\right].
\label{piauxfinal}
\end{split}
\end{equation} 

Using Eq. \eqref{freeaux} and Eq. \eqref{piauxfinal} we obtain the full Auxiliary-field propagator. We consider that the generated mass is much smaller than the external momentum, i.e, $p^2\gg \sigma^2_0$. Therefore, the propagator reads
\begin{equation}
\Delta_{\sigma}(p)=\frac{4 v_F^{2}}{\sqrt{p_0^{2}+v_F^{2}\textbf{p}^2}}.
\label{auxfull2}
\end{equation}

\subsection{The fermion self-energy}

\begin{figure}[H]
\centering
\includegraphics[scale=0.55]{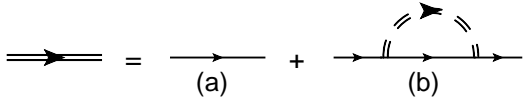}
\caption{\textit{The full fermion propagator}. $ (a) $ represents the free fermion propagator, $(b)$ 1-loop correction due to the Gross-Neveu interaction, within the auxiliary-field approach, where the double dashed line represents the full auxiliary-field propagator.}
\label{fermifull}
\end{figure}

The fermion propagator with the self-energy corrections, in the dominant order of 1/N, is shown in Fig. \ref{fermifull}.

This self-energy, due to the interactions with the auxiliary field,  is given by
\begin{equation}
\Sigma_{\sigma}(p)=\frac{1}{N} \int \frac{d^3 k}{(2 \pi)^3} S_F(p-k)\Delta_{\sigma}(k). \label{sss}
\end{equation}
Note that Eq. \eqref{sss} has a similar structure in comparison with Eq. \eqref{sigmaa2}. Therefore, we follow the same steps as before, yielding the self-energy, namely,
\begin{equation}
\Sigma_{\sigma}(p)= \Sigma_{\sigma}^{(0)}+ \gamma^0p_0\Sigma_{\sigma}^{(1a)} +v_F \gamma^ip_i \Sigma_{\sigma}^{(1b)},
\end{equation}
where $\Sigma_{\sigma}^{(0)}$, $\Sigma_{\sigma}^{(1a)}$, and $\Sigma_{\sigma}^{(1b)}$ are given by
\begin{equation}
\Sigma_{\sigma}^{(0)}=\frac{4v_F^{2}}{N}\int \frac{d^3 k}{(2 \pi)^3} \frac{\sigma_0}{k_0^{2}+ v_F^{2}\textbf{k}^2+\sigma_0^{2}}\frac{1}{\sqrt{k_0^{2}+ v_F^{2}\textbf{k}^2}},
\end{equation}
\begin{equation}
\Sigma_{\sigma}^{(1a)}=\frac{4v_F^{2}}{N}\int \frac{d^3 k}{(2 \pi)^3} \frac{v_F^{2}\textbf{k}^2- k_0^{2}+\sigma_0^{2}}{(k_0^{2}+ v_F^{2}\textbf{k}^2+\sigma_0^{2})^2}\frac{1}{\sqrt{k_0^{2}+ v_F^{2}\textbf{k}^2}}, 
\label{sigma1a}
\end{equation}
and
\begin{equation}
\Sigma_{\sigma}^{(1b)}=\frac{4v_F^{2}}{N}\int \frac{d^3 k}{(2 \pi)^3} \frac{k_0^{2}+\sigma_0^{2}}{(k_0^{2}+ v_F^{2}\textbf{k}^2+\sigma_0^{2})^2}\frac{1}{\sqrt{k_0^{2}+ v_F^{2}\textbf{k}^2}}. 
\label{sigma1b}
\end{equation}

Using the same variable change that has been made for obtain the Eq. \eqref{autofoton}, we obtain 
\begin{equation}
\begin{split}
\Sigma_{\sigma}(p,\sigma_0)&=\frac{2}{3\pi^2 N}\left(\gamma^0p_0+v_F\gamma^ip_i+3\sigma_0\right)\ln\left(\frac{\Lambda}{\Lambda_0}\right) \\ &+ \textrm{FT}.
\end{split}
\label{autoauxiliar}
\end{equation}

Note that the divergent part of Eqs. \eqref{sigma1a} and \eqref{sigma1b} are equal because of rotational symmetry, and in view of this, only the wave function renormalization is sufficient to renormalize the two points Green function due to the GN interaction. As a result, GN interaction does not change the renormalization of the Fermi velocity, as we shall see in the next section explicitly.

\subsection{Renormalization group function due GN and electromagnetic interations}

Since $\Pi^{\mu\nu}$ and $\Pi_{\sigma}$ are finite, within the dimensional regularization, we may conclude that $\gamma_{\sigma}=\gamma_{A_{\mu}}=0$, and, therefore, $\beta_e=0$. Hence, the RG equation reads  
\begin{equation}
\!\!\left[\Lambda \frac{\partial}{\partial \Lambda}+\beta_{v_F}\frac{\partial}{\partial v_F}+\beta_{\sigma_0}\frac{\partial}{\partial \sigma_0}- N_F \gamma_F\right]\Gamma^{(N_F,N_A,N_{\sigma})}(p_i)=0,
\label{GR}
\end{equation}
where $\Gamma^{(N_F,N_A,N_{\sigma})}(p_i=p_1,...,p_N)$ means the renormalized vertex functions. $N_F$, $N_A$, and $N_{\sigma}$ are the number of external lines of fermion, Gauge, and sigma fields, respectively. $\beta_{v_F}$, $\beta_{\sigma_0}$ and $\gamma_F$ are defined similarly as in section III. For our purpose, it is sufficient to consider only the vertex function for the fermion, i.e, $\Gamma^{(2,0,0)}$. Therefore, we can write 
\begin{equation}
\Gamma^{(2,0,0)}= \left(\gamma^0 p_0 + v_F\gamma^i p_i - \sigma_0\right) + \Sigma_{A_{\mu}}(p) + \Sigma_{\sigma}(p). \label{gammavertex}
\end{equation}

Using Eq. \eqref{gammavertex} into Eq. \eqref{GR} and using that $\beta_a=N^0\,\beta_a^{(0)}+\frac{1}{N}\,\beta_a^{(1)}+...$ for $a=v_F,\sigma_0$, and $\gamma_F=N^0\,\gamma_F^{(0)}+\frac{1}{N}\,\gamma_F^{(1)}+...$, we obtain, after some calculations, 
\begin{eqnarray}
\gamma_F&=&-\frac{2}{\pi^2 N}\left[2+\frac{2-\lambda^2}{\lambda\sqrt{1-\lambda^2}}\cos^{-1}(\lambda)\right] \nonumber\\ 
&+&\frac{2}{\pi N}\frac{1}{\lambda}+\frac{1}{3 \pi^2 N}, \label{anomala}
\end{eqnarray}
\begin{equation}
\beta_{v_{F}}=-\frac{4}{\pi^2 N}v_F\left[1+\frac{\cos^{-1}(\lambda)}{\lambda\sqrt{1-\lambda^2}}\right]+\frac{2}{\pi N}\frac{v_F}{\lambda}, \label{betavelocidad}
\end{equation}
and
\begin{equation}
\beta_{\sigma_0}=-\frac{2}{\pi^2 N}\sigma_0\left[4 +\frac{4 \cos^{-1}(\lambda)}{\lambda\sqrt{1-\lambda^2}}-\frac{2\pi}{\lambda}\right]+\frac{8 \sigma_0}{3 \pi^2 N}.
\label{betamass}
\end{equation}

We may conclude, from Eqs. \eqref{anomala}-\eqref{betamass}, that the Gross-Neveu interaction modifies both the anomalous dimension of the fermion as well as the beta function of the mass. On the other hand, \textit{it does not modify the beta function of the Fermi velocity}, hence, the Fermi velocity renormalization is insensitive to this interaction. The results in Eq. \eqref{betamass0} and Eq. \eqref{anomala0} are obtained by neglecting the last term in the rhs of Eq. \eqref{betamass} and Eq. \eqref{anomala}, respectively. These terms are generated by the four-fermion interaction. Other perturbative approach, considering this interaction, reveal similar conclusions, i.e, indeed, the Gross-Neveu interaction does not change the renormalization of $v_F$ \cite{fermisys}.

\subsection{The Critical Coupling Constant $\lambda_c$}
\begin{figure}[H]
\centering
\includegraphics[scale=0.65]{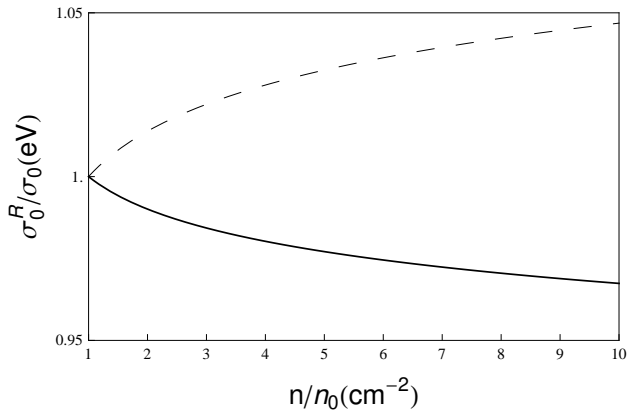}
\caption{\textit{Effects of Gross-Neveu interactions in the renormalization of the band gap}. The dashed line is the plot of Eq.~(\ref{sigR}) with $\lambda<\lambda_c\approx 0.66$ and $\lambda=0.4$. The common line is the plot of Eq.~(\ref{sigR}) with $\lambda>\lambda_c\approx 0.66$ and $\lambda=0.9$. } \label{comparison2}
\end{figure}

From Eq. \eqref{betamass} we may calculate the renormalized mass $\sigma^R_0$ as a function of the energy scale $\Lambda$. After a simple calculation, and performing $\Lambda/\Lambda_0\rightarrow (n/n_0)^{1/2}$, we find
\begin{equation}
\sigma^R_0(n)=\sigma_0\left(\frac{n}{n_0}\right)^{C^{{\rm GN}}_{\lambda}/2}, \label{sigR}
\end{equation}
where $\sigma_0\equiv \sigma(n_0)$ and
\begin{equation}
C^{{\rm GN}}_\lambda=-\frac{2}{\pi^2 N}\left[4+\frac{4 \cos^{-1}(\lambda)}{\lambda\sqrt{1-\lambda^2}}-\frac{2\pi}{\lambda}\right]+\frac{8}{3\pi^2 N} \label{Cl}
\end{equation}
is a known constant fixed by the coupling constant $\lambda$ and $N=2$. After solving $C^{{\rm GN}}_\lambda=0$ for $\lambda$, we find our critical value $\lambda_c\approx 0.66$. In this case, $\lambda=\lambda_c$, the value of the mass is the same for any energy scale. On the other hand, for $\lambda\neq \lambda_c$, the asymptotic behavior of $\sigma$ is dependent on the sign of $C^{{\rm GN}}_\lambda$. Indeed, for $\lambda>\lambda_c$, hence, $C^{{\rm GN}}_\lambda<0$, and $\sigma\rightarrow 0$ as $\Lambda$ goes to infinity. For $\lambda<\lambda_c$, we have $C^{{\rm GN}}_\lambda>0$ and $\sigma\rightarrow \infty$ as $\Lambda$ diverges. These last two cases are a consequence of the competition between PQED and the GN interaction, because the second term in the rhs of Eq.~(\ref{Cl}) is generated \textit{only due} to the GN interaction, being the theoretical prediction with the presence of a considerable disorder. In Fig.~\ref{comparison2}, we summarize these different asymptotic regimes.

From Fig.~\ref{comparison2}, we conclude that the behavior of $m(n)$ remains the same for $\sigma^R_0(n)$ whether $\lambda>\lambda_c$. The behavior of $m(n)$ has been compared with experimental data of both WSe$_2$ and MoS$_2$ in Sec.~II. Nevertheless, considering that $\lambda=\pi \alpha/4$, and $\alpha=e^2/4\pi \epsilon v_F$, we also conclude that one would decrease the value of $\lambda$ whether the factor $\epsilon v_F$ increases. This could be obtained either by choosing a proper substrate (with large $\epsilon$) or by increasing the value of $v_F$ which naturally occurs for clean samples \cite{geim}. A fine tunning of $\lambda$ close to $\lambda_c$ yields a case where the band gap remains the same at any energy scale. We believe that these theoretical predictions could be useful for applications in several two-dimensional materials (whom are described by the massive Dirac equations at low energies), in particular, for studying electric-field tunning of band energies \cite{WSe2gap}.

\section{Conclusions} 

In this work, we investigate the renormalization of the bandgap, in both WSe$_2$ and MoS$_2$.  Since these materials have $sp^2$ hybridization, the electromagnetic interaction between the massive quasiparticles of these systems can be described by Pseudo quantum electrodynamics. Using the renormalization group approach, in the dominant order in $1/N$, we show that our results are in excellent agreement with recent experimental measurements of the bandgap in these materials. A more realistic model for describing transport properties in these systems should include four-fermions interactions, once this interaction could simulate a disorder/impurity-like microscopic interaction. Thus, we also investigate the influence of a GN-type interaction in the behavior of the renormalization group function of PQED, where initially massless fermions acquire mass by chiral symmetry breaking. We show that the presence of a GN-type interaction does not change the behavior of the renormalized Fermi velocity, which is well-known in the literature. On
the other hand, the mass function has a richer behavior, which allows us to recognize a single fixed point at
$\lambda_c \approx 0.66$, representing an ultraviolet fixed point. This
renormalized mass shows a different behaviors whether $\lambda$ is above or below the critical point. This result could
be relevant for applications of 2D materials that rely on
tunable band gaps.  

We hope that our result clarify the relevance and beauty of applications of quantum field theory in the description of electrons in 2D materials. Because the fine-structure constant may be increased, it would be relevant to understand the nonperturbative effects on the renormalization of $m$ as well as to calculate the higher-temperature effects. We shall investigate this elsewhere. 

\section*{Acknowledgments}

L. F.  is partially supported by Coordena\c{c}\~{a}o de Aperfei\c{c}oamento de Pessoal de N\'{i}vel Superior – Brasil (CAPES), finance code 001; V. S. A. and L. O. N. are partially supported by Conselho Nacional de Desenvolvimento Cient\'{i}fico e Tecnol\'{o}gico (CNPq) and by CAPES/NUFFIC, finance code 0112. F. P. acknowledge the financial support from DIUFRO Grant DI18-0059 of the Direcci\'{o}n de Investigaci\'{o}n y Desarrollo, Universidad de La Frontera. F. P. acknowledge the hospitality of the Faculdade de F\'{i}sica of the Universidade Federal do Par\'{a}, Bel\'{e}m where part of this work was done. The work of E.C.M. is supported in part by CNPq and FAPERJ. The authors would like to thank J. Gonz\'{a}lez for his comments in the  Ref. \cite{Gui}.

\section*{APPENDIX: SOME DETAILS OF THE
CALCULATIONS}

\subsection*{Appendix A: Gauge-self Energy}
In this appendix we derive Eq. \eqref{polk}, i.e, 
\begin{equation}
\Pi^{\mu \nu}(p)= -N\, Tr \int \frac{d^3 k}{(2\pi)^3} \Gamma^{\mu} S_F(k)\Gamma^{\nu}S_F(k+p),
\end{equation}   
where the interaction vertex should be understood as $\Gamma^{\mu}\rightarrow \gamma^0 e/\sqrt{N}$ due to the static approximation. where the trace operation is understood on Lorentz indices and internal symmetry. Using matrix representation $\gamma$ as $4\times 4$, we have the following trace properties 
\begin{align}
{\rm Tr}\left[\gamma^{\mu}\gamma^{\nu}\right] &= -4\delta^{\mu \nu},\nonumber \\ 
{\rm Tr}\left[\gamma^{\mu}\gamma^{\alpha}\gamma^{\nu}\right] &= 0,\label{tr}
 \\ \nonumber
{\rm Tr}\left[ \gamma^{\mu}\gamma^{\alpha}\gamma^{\nu}\gamma^{\beta}\right] & = 4\left( \delta^{\mu \alpha}\delta^{\nu \beta}-\delta^{\mu \nu}\delta^{\alpha \beta} + \delta^{\mu \beta}\delta^{\nu \alpha}\right).
\end{align} 
Performing the trace operations and using the Feynman parametrization, we have
\begin{equation}
\Pi^{00}(p)=-4e^2\int_0^{1}dx\int \frac{d^3 k}{(2\pi)^{3}} \frac{{\rm Num}}{\left[{\rm Den}\right]^2},
\end{equation}
where ${\rm Num}$= $k_0(k_0+p_0)-\delta^{ij}v_F^{2}k_i(k+p)_j-m^{2}$ and ${\rm Den}$ = $(k_0+xp_0)^2+x(1-x)p_0^{2} + v_F^{2}(\textbf{k}+x\textbf{p})^2+x(1-x) v_F^{2}\textbf{p}^2+m^{2}$. Solving the integrals over $k_0$ and $\textbf{k}$, we obtain 
\begin{equation}
\begin{split}
\Pi^{00}(p)&=-\frac{e^2 \mu^{2 \epsilon}}{2 \pi v_F} \int_0^{1}\!\!\!\! dx \left\lbrace \sqrt{\Delta_2} - \frac{x(1-x)(p_0^{2}-v_F^{2}\textbf{p}^2)}{v_F^{2} \sqrt{\Delta_2}}\right. \\ &\left.+\frac{m^{2}}{v_F^{2} \sqrt{\Delta_2}} \right\rbrace,
\end{split}
\end{equation} 
where $\Delta_2$=$\frac{1}{v_F^{2}} \left[ x(1-x)(p_0^{2}+v_F^{2}\textbf{p}^2)+m^{2} \right]$. Therefore, after of integration in the Feynman parameter, the time-component of the polarization tensor, in the small-mass limit $p^2\gg m^2$, is 
 \begin{equation}
\Pi^{00}=-\frac{e^2 }{ 8}\left[\frac{\textbf{p}^2}{\sqrt{p_0^{2}+v_F^{2}\textbf{p}^{2}}}-\frac{4\textbf{p}^2 m^{2}}{(p_0^{2}+v_F^{2}\textbf{p}^{2})^{\frac{3}{2}}}\right].
\end{equation}
 
 \section*{Appendix B: Loop Integral Calculation}

The integrals we solve in our model have a particular feature due to the Lorentz symmetry breaking. For clarifying this point, we show how to obtain the function $f_0(\lambda)$, given by Eq. \eqref{f0}. First, we made a variable change $v_F k_i$ $\rightarrow$ $\bar{k}_i$.Thereafter, for solving the integrals, we use spherical coordinates, hence,
\begin{align*}
k_0 &= k \cos\theta,\\
|\overline{\vec{k}}|&=k \sin\theta,\\
d^3\overline{k}&=k^2 \sin\theta\, dk \, d\theta \, d \phi.
\end{align*}
Therefore, we write Eq. \eqref{f0} as
 \begin{equation}
\begin{split}
\Sigma_{A_{\mu}}^{(0)}&=-\frac{e^2 \sigma_0}{2(2\pi)^2\epsilon N v_F} \int_0^{\Lambda}dk \frac{k}{k^2+\sigma_0^{2}}\\ &\times\int_0^{\pi}d\theta\frac{1}{1+\lambda(1-4\frac{\sigma_0^{2}}{k^2})\sin\theta},
\end{split}
\end{equation}
where the term $\sigma_0^{2}/k^2 \approx 0$, thereby,
\begin{equation}
\int_0^{\pi}d\theta\frac{1}{1+\lambda \sin \theta}= \frac{2\cos^{-1}(\lambda)}{\sqrt{1-\lambda^2}}. \label{presul}
\end{equation}
Because Eq.~(\ref{presul}) does not depend on the momentum of the loop, we can define it as $f_0(\lambda)$ for $\lambda < 1$. The other functions, $f_1(\lambda)$ and $f_2(\lambda)$, are similarly obtained from the angular integrals in Eq.~\eqref{f1} and Eq.~\eqref{f2}.

\subsection*{Appendix C: Auxiliary-self Energy}
We start by calculating the auxiliary-self energy, within Eq. \eqref{Dsigma}, given by 
\begin{equation}
\begin{split}
\Pi_{\sigma}(p)&= -\left(-\frac{1}{\sqrt{N}}\right)^2 N\,Tr \int \frac{d^3 k}{(2 \pi)^{3}} S_F(k)S_F(p+k), \label{auxself0}
\end{split}
\end{equation}
where the trace operation given the Eq. \eqref{tr}, we use the dimensional regularization scheme for calculating the linear divergence of Eq.~(\ref{auxself0}). Hence,
\begin{equation}
\Pi_{\sigma}(p)=4\int_0^{1} dx \int \frac{d^3 k}{(2 \pi)^{3}} \frac{{\rm Num_2}}{\left[{\rm Den}\right]^2},
\end{equation}
where ${\rm Num_2}$=$k_0(k_0+p_0) + v_F^{2}\delta^{ij}k_i(k+p)_j -\sigma_0^{2}$ and ${\rm Den}$ = $(k_0+xp_0)^2+x(1-x)p_0^{2} + v_F^{2}(\textbf{k}+x\textbf{p})^2+x(1-x) v_F^{2}\textbf{p}^2+\sigma_0^{2}$. Solving the integrals over $ k_0 $ and $\textbf{k}$, we obtain
\begin{equation}
\Pi_{\sigma}(p)=-\frac{2\mu^{\epsilon}}{\pi v_F^{2}} \int_0^{1}dx \sqrt{x(1-x)(p_0^{2}+v_F^{2}\bf{p}^{2})+\sigma_0^{2}}, \label{solvingx}
\end{equation}
where $x$ is a Feynman parameter. Integration over $x$ in Eq.~(\ref{solvingx}) yields Eq.~(\ref{piauxfinal}).

\hspace{0.5cm}

\bibliographystyle{elsarticle-num}
\bibliography{<your-bib-database>}

\end{document}